\documentclass[12pt]{article}
\title{Ten-dimensional super-Yang-Mills with nine off-shell supersymmetries}
\usepackage{mathrsfs}
\usepackage{amsmath}

\global\arraycolsep=1pt
\usepackage[colorlinks=true,backref=true,linkcolor=black,anchorcolor=black,citecolor=black,filecolor=black,menucolor=black,pagecolor=black,urlcolor=black]{hyperref}
\usepackage[Symbol]{upgreek}
\usepackage{bbm}
\usepackage{dsfont}
\usepackage{amssymb}
\usepackage{textcomp}

\newcommand{\Scal}[1]{\Bigl ({#1} \Bigr )}

\def\bea{\begin{eqnarray}}
\def\eea{\end{eqnarray}}
\def\be{\begin{equation}}
\def\ee{\end{equation}}
\newcommand{\CR}{\nonumber \\*}
\newcommand{\del}{\partial}

\newcommand{\trace}{\hbox {Tr}~}

\def\L{{\cal L}}

\DeclareMathAlphabet{\mathpzc}{OT1}{pzc}{m}{it}

\def\d{\mathfrak{d}}
\def\e{\mathfrak{e}}

\def\t{\tilde}

\usepackage[colorlinks=true,backref=true,linkcolor=black,anchorcolor=black,citecolor=black,filecolor=black,menucolor=black,pagecolor=black,urlcolor=black]{hyperref}

\usepackage[Symbol]{upgreek}
\usepackage{caption}
\usepackage{bbm}
\usepackage{dsfont}
\usepackage{amssymb}
\usepackage{textcomp}
\def\bea{\begin{eqnarray}}
\def\eea{\end{eqnarray}}
\def\be{\begin{equation}}
\def\ee{\end{equation}}

\def\L{{\cal L}}

\def\t{\tilde}

\def\N{\mathcal{N}}

\def\lag{\mathcal{L}}
\def\d{\delta}
\def\e{\epsilon}
\def\ve{\varepsilon}
\def\Ga{\Gamma}

\def\hGa{\hat{\Gamma}}
\def\t{\theta}
\def\z{\zeta}
\def\l{\lambda}

\def\susy {\delta^{\scriptstyle \mathpzc{Susy}}}
\def\gauge{\delta^{\scriptstyle  \rm gauge}}

\def\be{\begin{equation}}
\def\ee{\end{equation}}
\def\bea{\begin{eqnarray}}
\def\eea{\end{eqnarray}}
\def\bdis{\begin{displaymath}}
\def\edis{\end{displaymath}}
\def\corr{$\clubsuit$}
\def\nn{\nonumber}

\begin{document}
\allowdisplaybreaks[1]
\renewcommand{\thefootnote}{\fnsymbol{footnote}}
\def\corr{$\spadesuit $}
\def\trefle{ $\clubsuit$}
\begin{titlepage}
\begin{flushright}
CERN-PH-TH/2006-047
\end{flushright}
\begin{flushright}
IFT-P.008/2007
\end{flushright}
\begin{center}
{{\Large \bf

Ten-dimensional super-Yang--Mills with nine off-shell supersymmetries

 }}
\lineskip .75em
\vskip 3em
\normalsize
{\large L. Baulieu\footnote{email address: baulieu@lpthe.jussieu.fr},
 N.~Berkovits\footnote{email address: nberkovi@ift.unesp.br},
 G. Bossard\footnote{email address: bossard@lpthe.jussieu.fr}, 
A. Martin\footnote{email address: alexis.martin@lpthe.jussieu.fr}\\
\vskip 1em
$^{* }${\it Theoretical Division CERN}\footnote{ CH-1211 Gen\`eve, 23, Switzerland
}
\\
$^{*\ddagger \mathsection}${\it LPTHE, CNRS and Universit\'es Paris VI - Paris VII}\footnote{
4 place Jussieu, F-75252 Paris Cedex 05, France.}
\\
$^{\dagger}$ {\it Instituto de F\'\i sica Te\'orica, State University of S\~ao Paulo}\footnote{
Rua Pamplona 145, 01405-900, S\~ao Paulo, SP, Brasil}
\\
$^{\ddagger}${\it Instituut voor Theoretische Fysica}\footnote{Valckenierstraat 65, 1018XE Amsterdam, The Netherlands}
}

\vskip 1 em
\end{center}
\vskip 1 em
\begin{abstract}

After adding 7 auxiliary scalars to the d=10 super-Yang--Mills action,
9 of the 16 supersymmetries close off-shell. In this paper, these 9
supersymmetry generators are related by dimensional reduction to scalar
and vector topological symmetry in $\N$=2 d=8 twisted super-Yang--Mills.
Furthermore, a gauge-invariant superspace action is constructed for d=10
super-Yang--Mills where the superfields depend on 9 anticommuting $\theta$
variables.

\end{abstract}

\end{titlepage}
\renewcommand{\thefootnote}{\arabic{footnote}}
\setcounter{footnote}{0}



\renewcommand{\thefootnote}{\arabic{footnote}}
\setcounter{footnote}{0}


 \def\stop{$\blacksquare$}

\section{Introduction}
The off-shell field content of $10$-dimensional
super-Yang--Mills theory has an excess of   seven fermionic degrees
of freedom as compared to the number of    gauge invariant bosonic
degrees of freedom.
To balance this mismatching, it was proposed in \cite{berko} to
add to the supersymmetry
transformation laws a set of seven auxiliary scalar
fields $G_a$, together
with a $\sum_a G_a^2$ term in the action. In order for the algebra to
close off-shell, the parameters associated with the supersymmetry
transformations must obey some
identities.
However, there is no linear solution to these identities, and 
thus no conventional supersymmetric formulation which permits the algebra
to completely close.

It has been demonstrated in \cite{berko} that it is impossible to 
construct more than nine consistent solutions of these identities. 
Thus,  only nine supersymmetry generators can generate an algebra that 
closes off-shell.  
These nine supersymmetry generators in ten-dimensional super-Yang--Mills
are related to the octonionic division algebra in the same manner that
the supersymmetry generators in three-, four-, and six-dimensional 
super-Yang--Mills are related to the real, complex, and quaternionic
division algebras. However, the non-associativity of octonions makes
the ten-dimensional supersymmetry algebra more complicated than in the 
other dimensions. 

On the other hand, the  $\N=2$  twisted $8$-dimensional 
super-Yang--Mills theory, which is a  particular dimensional  reduction 
of the   $10$-dimensional theory,  has been 
determined  in  \cite{BBT}  by the invariance under a subalgebra of the maximal 
Yang--Mills supersymmetry. This subalgebra    
is small enough to close independently of equations of motion  
with a finite set of auxiliary fields, and yet is large enough to determine 
the Yang--Mills supersymmetric theory. It  is   also made of    nine 
generators. The latter  can be geometrically understood and constructed as
scalar and vector topological Yang--Mills symmetries. This   $8$-dimensional 
topological  symmetry can be built independently of the notion of 
supersymmetry, but, surprisingly, the latter   symmetry with 16 generators   
can be fully  recovered at the end of the construction.

  The aim of this paper is to make a bridge between the results of 
\cite{berko} and \cite{BBT}.
We will find that in $10$-dimensional flat space  with Lorentz group 
$SO(1,9)$ reduced to $SO(1,1)\times Spin(7)$, the supersymmetry algebra can be twisted 
such that the $10$-dimensional super-Yang--Mills theory is determined  by a
supersymmetry algebra with 9 generators, which   is related
by   dimensional  reduction to the twisted $\N=2$  $8$-dimensional 
super-Yang--Mills theory. Reciprocally, the extended curvature equation of the 
$\N=2$ 8-dimensional supersymmetric theory   can be ``oxidized"  into  
an analogous 10-dimensional equation that determines the 
supersymmetry algebra and    $10$-dimensional super-Yang--Mills action.  
We argue that the largest
symmetry group that can preserve an off-shell subalgebra of supersymmetry is 
$SO(1,1)\times Spin(7)$, and we obtain the most general $SO(1,1)\times Spin(7)$ covariant solution 
of the identities defined in \cite{berko}. The supersymmetry algebra that 
we derive is exactly the one obtained by the twist operation.

We then define a superspace involving nine Grassmann $\theta$
variables such that the off-shell 
supersymmetry subalgebra acts in a manifest way on the super-Yang--Mills 
superfields. Using these off-shell superfields, a superspace action is 
constructed which reproduces the ten-dimensional super-Yang--Mills action 
including the seven auxiliary scalar fields $G_a$. Although this
superspace action is manifestly invariant under only a $Spin (7)\times
SO(1,1)$ 
subgroup of $SO(9,1)$, it is manifestly invariant under nine supersymmetries
as well as
gauge transformations. This can be compared with the light-cone superspace
action for ten-dimensional super-Yang--Mills which is manifestly invariant
under eight supersymmetries and
an $SO(8)\times SO(1,1)$ (or $U(4)\times SO(1,1)$) subgroup of
$SO(9,1)$, but is not manifestly
invariant under gauge transformations.

\section{Ten dimensional supersymmetric Yang--Mills with auxiliary fields}

The   Poincar\'e supersymmetric Yang--Mills theory in ten
dimensional Minkowski space
 contains a gauge field $A_\mu$ ($\mu=1,\cdots 10$) and a
sixteen-component Majorana--Weyl spinor $\Psi$, with values in the
Lie algebra of some gauge group.  
In order to balance the gauge-invariant off-shell degrees
of freedom, one  can   introduce a set of scalar fields $G_a$
($a=1,\cdots 7)$ which count for the $7$ missing bosonic degrees of
freedom  \cite{berko}. The Lagrangian is given by
\be \label{lag 10D}
\lag=Tr\{-{1\over4 }
F_{\mu\nu}F^{\mu\nu}+\frac{i}{2}(\bar{\Psi}\hat{\Ga}^\mu
D_\mu\Psi)+8 G_a G_a\}
\ee
where $\hat\Ga^\mu$ are the ten-dimensional gamma matrices.
As shown in \cite{berko}, the  action (\ref {lag 10D}) is
invariant under the following supersymmetry transformations, which
depend on the ordinary Majorana--Weyl  parameter $\e$ and on seven
other spinor parameters $v_a$
\begin{eqnarray} \label{lois 10D}
\d A_\mu &=& i\bar{\e} \hat{\Ga}_\mu \Psi \nn\\
\d \Psi &=& \hat{\Ga}^{\mu\nu}F_{\mu\nu}\e+4G_a v_a \\
\d G_a &=& -\frac{i}{4}\bar{v}_a \hat{\Ga}^\mu D_\mu \Psi
\end{eqnarray}
The  commuting spinor  parameters  $v_a$ must be  constrained   as
follows
\be \label{cond}
\bar{v}_a\hat{\Ga}_\mu\e = \bar{v}_a\   \hat{\Ga}_\mu v_b  -
\d_{ab}\bar{\e} \hat{\Ga}_\mu\e=0
\ee
The  transformations  (\ref {lois 10D}) generate a closed algebra
modulo gauge transformations and equations of motion
\be \label{alg 10D}
\{\d,\hat{\d}\} \approx
-2i\bar{\e}\hat{\Ga}^\mu\hat{\e}\, \partial_\mu - 2 i \gauge ( \bar{\e}\hat{\Ga}^\mu A_\mu \hat{\e} )
\ee
and  close  independently of
equations of motion when
\bea\label{ber}
(\hat{\e},\hat{v_a})
\eea
is some linear combination of
$(\hat{\Ga}^{\mu\nu}\e,\hat{\Ga}^{\mu\nu}v_a)$. To recover
conventional supersymmetry transformations, one must have   a
solution for  $v$  in  (\ref{cond}) that is linear in $\e$. This
in turn will give a realisation of (\ref{alg 10D}) which, thanks
to (\ref{ber}), will effectively hold off-shell.

Using  octonionic notations
and  light-cone coordinates, a solution was found   for  the $v$'s
and $\e$ in  \cite{berko}   that  preserves nine supersymmetries. This 
solution is only covariant under $SO(1,1) \times Spin(7) \subset SO(1,9)$. 
In fact, in order to define the $v$'s as linear combinations of  $\e$, we 
must reduce the covariance to a subgroup $ H$ that admits a 7-dimensional 
representation. Moreover, since the maximal sub-algebra that can be closed 
off-shell contains 9 supersymmetry generators, the Majorana--Weyl spinor 
representation of $Spin(1,9)$ must decompose into ${\bf 7+ 9}$ of $H$. 
The biggest subgroup of $ SO(1,9)$ that satisfies these criteria is 
$SO(1,1) \times Spin(7)$.

\subsection{Light-cone variables}
The choice of light-cone variables implies a reduction of the
Lorentz group as
\be
SO(1,9) \to SO(8)\times SO(1,1)
\ee
where the spinor $\Psi\in
\mathbf{16_+}$ of $SO(1,9)$ decomposes into one chiral and one
antichiral spinor of $Spin(8)$,   $\Psi \to \lambda_1\oplus\lambda_2 \in \mathbf{8_+^{-1}}\oplus
\mathbf{8_-^1}$, as well as $\e\to\e_1\oplus\e_2$ and 
$v_a\to v_{a1}\oplus v_{a2}$. The connection $A_\mu\in \mathbf{10}$ of
$SO(1,9)$ decomposes according to $A_\mu\to A_i\oplus A_+\oplus
A_-\in \mathbf{8_v^0}\oplus \mathbf{1^2}\oplus \mathbf{1^{-2}}$ of
$SO(8)\times SO(1,1)$, where the superscripts denote the
eigenvalue associated with the $SO(1,1)$ factor and
$A_{\pm}=A_0\pm A_9$.

We can  consider a gamma matrix algebra of $Cl(1,9)$ in terms of
gamma matrices of $Cl(0,8)$
\bea
\hGa_0 &=& (i\sigma_2)\otimes\Ga_9\nn\\
\hGa_i &=& \sigma_2\otimes\Ga_i \qquad(i=1\dots 8)\nn\\
\hGa_9 &=& \sigma_1\otimes 1
\eea
and $\hGa_{11} \Psi= \sigma_3 \otimes 1 \, \Psi= \Psi $. These matrices obey
$[\hGa_\mu,\hGa_\nu]=2\,\eta_{\mu\nu}$, with the metric
$\eta_{\mu\nu}=$
$\textrm{diag}(-1,+1,\dots,+1)$ and
$[\Ga_i,\Ga_j]=2\,\d_{ij}$. The decomposition $SO(1,9)\to
SO(8)\times SO(1,1)$ is performed by taking $A_\pm=A_0\pm A_9$ and
by projecting $\Psi$ to   $\hGa_+\Psi \to \l_2$ and $\hGa_-\Psi \to
\l_1$, with $\Ga_9\l_1=\l_1$ and $\Ga_9\l_2=-\l_2$.

The transformations laws (\ref{lois 10D})   read
\bea \label{lois 8D}
\d A_i &=& -i\bar{\e} \Ga_i \l \nn\\
\d A_+ &=& -2\bar{\e}_2\l_2\nn\\
\d A_- &=& 2\bar{\e}_1\l_1\nn\\
\d \l_1 &=&
(F_{ij}\Ga_{ij}+\frac{1}{2}F_{+-})\e_1+iF_{i-}\Ga_i\e_2+4G_a
v_{a1}\nn\\
\d \l_2 &=&
(F_{ij}\Ga_{ij}-\frac{1}{2}F_{+-})\e_2-iF_{i+}\Ga_i\e_1+4G_a
v_{a2}\nn\\
\d G_a
&=&\frac{i}{4}\bar{v}_a\Ga_iD_i\l+\frac{1}{4}\bar{v}_{a1}D_+\l_1-\frac{1}{4}\bar{v}_{a2}D_-\l_2
\eea
together with the constraints
\bea \label{light-cone cond}
&\bar{v}_a\Ga_i\e =\bar{v}_{a1}\e_1=\bar{v}_{a2}\e_2=0&\\ \nn
\bar{v}_{a1} v_{b1} = \delta_{ab} \bar{\e}_1\e_1\quad &\bar{v}_a\Ga_i v_b 
= \delta_{ab} \bar{\e}\Ga_i\e&\quad \bar{v}_{a2} v_{b2} = 
\delta_{ab} \bar{\e}_2\e_2
\eea
The Lagrangian is given by
\bea
\lag=Tr\Big(-\frac{1}{4}(F^{ij})(F_{ij})-\frac{1}{4}(F^{i+})(F_{i+})-\frac{1}{4}(F^{i-})(F_{i-})-\frac{1}{8}(F^{+-})(F_{+-})\nn\\
-\frac{i}{2}\bar{\l}\Ga_iD_i\l-\frac{1}{2}\bar{\l}_1D_+\l_1+\frac{1}{2}\bar{\l}_2D_-\l_2+8(G_a)^2\Big).
\eea

\subsection{Twisted variables}
In \cite{berko}, the light-cone projections $\e_1$ and $\e_2$ of the spinor parameter are expressed in terms of octonions and the imaginary components of $\e_1$ are set to zero, that is $\e_1$ is taken real. Formally, the reality constraint on the supersymmetry parameter
$\e_1$ implies the decomposition of the corresponding representation $\mathbf{8_+}\to \mathbf{1}\oplus
\mathbf{7}$ associated to the inclusion $Spin(7) \subset 
Spin(8)$\footnote{A discussion of various solutions of
(\ref{cond}) can be found in \cite{Evans} together with their
invariance groups. In particular, a solution is presented
preserving nine supersymmetries, but with a reduction of $SO(1,9)
\to G_2\times SO(1,1)$.
}

\bea
\e_1 \in \mathbf{8_+} &\to&
\mathbf{1}\oplus \mathbf{7}\nn\\
 \e_2\in \mathbf{8_-}  &\to&  \ \ \, \mathbf{8}
\eea
where $\mathbf{1}$,$\mathbf{7}$ and $\mathbf{8}$ define the
scalar, vectorial respectively spinorial representations of
$Spin(7)$. The reality constraint is then equivalent to
retaining just the singlet part of $\e_1$, isolating $9$
supersymmetries.

For expressing the  decomposition $SO(8)\to Spin(7)$,
 it is   convenient  to
introduce    projectors onto the irreducible representations of
$Spin(7)$. In order to do so, we use the spinor $\z$ scalar of  $Spin(7)$. We take it
chiral and of norm $1$ and define \be \label{octo_4-form}
(4!)\bar{\z}\Ga_{ijkl}\z\equiv
\Omega_{ijkl}\quad\textrm{with}\quad \bar{\z}\z=1,\\
\ee
where $\Omega_{ijkl}$ stands for the octonionic $Spin(7)$
invariant $4$-form. It can be used to construct orthogonal
projectors to  decompose   the adjoint
representation $\mathbf{28}$ of $Spin(8)$ into the irreducible $Spin(7)$ ones $\mathbf{21}\oplus
\mathbf{7}$
\bea
P_{ij}^{+\phantom{1}
kl}&\equiv&\frac{3}{4}(\d_{\,ij}^{\,kl}+\frac{1}{6}\Omega_{\,ij}^{\,kl})\\
\nn P_{ij}^{-\phantom{1}
kl}&\equiv&\frac{1}{4}(\d_{\,ij}^{\,kl}-\frac{1}{2}\Omega_{\,ij}^{\,kl})
\eea
The supersymmetry parameter can then be expressed as
\bea \label{split epsilon}
\e_1 &=& \bar{\omega}\z
+\Ga_{ij}\nu^{ij}\z\nn \\
\e_2 &=& i\Ga_i \varepsilon^i\z
\eea
where $\nu_{ij}=P_{ij}^{-\phantom{1} kl}\nu_{kl}$. This provides us with  a
decomposition of the supersymmetry generators as
 an antiselfdual tensorial charge $\delta_{ij}$,    a
scalar charge $\delta_0$ and a vectorial charge $\delta_i$, of which we will retain just the scalar and vectorial ones.

\subsection{Decoupling and resolution of the constraints}

In terms of $Spin(7)$ representations, the relations (\ref{light-cone cond}) together with the explicit expression for the supersymmetry paramaters (\ref{split epsilon}) read
\bea \label{twisted cond}
\bar{v}_1^{ij}(\bar{\omega}+\nu^{kl}\Ga_{kl})\z=0,\quad&\bar{v}_1^{ij}\Ga_m
i\ve^k\Ga_k\z+\bar{v}_2^{ij}\Ga_m
(\bar{\omega}+\nu^{kl}\Ga_{kl})\z
=0,&\quad\bar{v}_2^{ij}i\ve^k\Ga_k\z=0\\ \nn \bar{v}_{1ij}
v_1^{kl} = 8P^{-\phantom{1}kl}_{
ij}(\bar{\omega}^2+2\vert\nu\vert^2), &\bar{v}_{ij}\Ga_m v^{kl} =
16iP^{-\phantom{1}kl}_{ ij} (\bar{\omega}\ve_m+4\ve^n\nu_{mn}), &
\bar{v}_{2ij} v_2^{kl} = -8P^{-\phantom{1}kl}_{ ij}
\vert\ve\vert^2 \eea In order to find the most general covariant
solution for $v_{ij}$ that is linear in the supersymmetry
parameters, we consider \bea
v_1^{ij} &=& a\bar{\omega}\Ga^{ij}\z+e \nu^{ij}\z+f\nu_{kl}\Ga^{ijkl}\z \nn\\
v_2^{ij} &=& b\ve^k\Ga_k\Ga^{ij}\z+c\ve^{[ i}\Ga^{j] -}\z
\eea
from which we can show, remembering 
that all terms, but $\bar{\z}\z$ and $\bar{\z}\Ga_{ijkl}\z$,
vanish, that (\ref{twisted cond}) is verified for 
the solution $\nu^{ij}=0$, $a=-ib=2$ and $c=0$, that is
\begin{displaymath}
\begin{array}{ll}
\e_1 = \bar{\omega}\z \qquad &  v_1^{ij} = 2\bar{\omega}\Ga^{ij}\z\\
\e_2 = i\Ga_i \varepsilon^i\z \qquad &  v_2^{ij} = 2i\ve^k\Ga_k\Ga^{ij}\z
\end{array}
\end{displaymath}
As expected, this solution provides us with a set of nine components
parameterized by $\bar\omega$ and $\varepsilon^i$, 
which form the maximal
set of supersymmetry generators that can generate an off-shell algebra.\\

The fields of the theory decompose according to
\bea
\l_1 \in \mathbf{8_+} &\to&
\mathbf{1}\oplus \mathbf{7}\nn
\\
\l_2\in \mathbf{8_-} ,\ A_i \in \mathbf{8_v} &\to&  \ \ \,
\mathbf{8}
\\ \nn
\eea
and $G_a$ is reexpressed in terms of $G_{ij}^- \in \mathbf{7}$ as
$G_{8a}=G_a$ and $G_{ab}={C_{ab}}^c G_c$, where ${C_{ab}}^c$ are the
structure constants of the imaginary octonions.
One has explicitly
\bea
\lambda_1 &=& \eta\z+\Ga_{ij}\chi^{ij}\z  \nn  \\
\lambda_2 &=& i\Ga_i \psi^i\z  \eea and
\bea
\eta &=& \bar{\z}\l_1 \nn\\
\chi_{ij} &=& -\frac{1}{2}\bar{\z}\Ga_{ij}\l_1   \nn  \\
\psi_i &=& -i\bar{\z}\Ga_i\l_2
\eea
where $\chi_{ij}=P_{ij}^{-\phantom{1} kl}\chi_{kl}$.
The supersymmetry transformations are now generated by
\be
\susy=\bar{\omega} \delta_0+\ve^i \delta_i
\ee
We display here the resulting transformation laws in a form which is more 
convenient with respect to the approach related to theories of cohomological type (  called BRSTQFTs  in  the terminology of  \cite{bakasi})   of the
next section. That is, we redefine $G_{ij}\rightarrow G_{ij}-
P_{ij}^{-kl}F_{kl}$ 
and redefine some of the fields by scale factors to get
\bea
\delta_0 A_i &=& \psi_i\nn\\
\delta_0A_+ &=& 0\nn\\
\delta_0A_- &=& \eta
\nn\\
\delta_0\psi_i &=& -F_{i+}\nn\\
\delta_0\eta &=& F_{+-}\nn\\
\delta_0\chi_{ij} &=& G_{ij}\nn\\
\delta_0G_{ij} &=& D_+\chi_{ij}
\\
 \nn\\
 \delta_i A_j &=& -\d_{ij}\eta -\chi_{ij}\nn\\
\delta_i A_+ &=& -\psi_i\nn \\
\delta_i A_- &=& 0 \nn\\
\delta_i\psi_j &=&
F_{ij}+G_{ij}+\d_{ij}F_{+-}\nn\\
\delta_i\eta &=& F_{i-} \nn\\
\delta_k\chi_{ij} &=& 8P^{-\,\,l}_{ijk}F_{l-}\nn \\
\delta_k G_{ij} &=& D_k\chi_{ij}-8P^{-\,\,l}_{ijk}(D_l\eta-D_-\psi_l)
\eea
The algebra closes independently of the equations of motion as
\bea
\delta_0^2 &=& \partial_+ + \gauge(A_+) \nn\\
\delta_{\{i} \delta_{j\}} &=& \d_{ij}(\partial_- + \gauge(A_-)) \nn\\
\{\delta_0,\delta_i\} &=& \partial_i + \gauge(A_i)
\eea
and the action becomes
\bea \label{10D action}
S = \int_M d^{10} x\, \trace
\biggl(\frac{1}{2}\,G^{ij}(F_{ij}+\frac{1}{4}G_{ij})-\chi^{ij}
(D_i\psi_j+\frac{1}{8}D_+\chi_{ij})+\eta
D_i\psi^i                         +\nn\\
+(F^i_{\phantom{i}-})(F_{i+})-\psi^i D_-\psi_i+ (F_{+-})^2- \eta
D_+\eta \biggr).
\eea

The formal dimensional reduction on the ``Minkowski torus'' consists trivially here to neglect the non-zero modes of the operators $\partial_\pm $. Doing so we
recover the eight-dimensional cohomological action and its twisted supersymmetry algebra, obtained in \cite{BBT}
by twisting the eight-dimensional theory. In the next section, we discuss this link to eight-dimensional
cohomological theory in more details.

\section{Link with  eight-dimensional Yang--Mills BRSTQFT}

In this section, we will directly obtain the light-cone twisted subalgebra of
$10$-dimensional super-Yang--Mills of the last section. It will 
 close off-shell by construction and will be
inspired by the  analogous known  subalgebra  of maximal
twisted\footnote{Here, the word twist means the mapping between
forms and spinors that is allowed reducing the $SO(8)$ covariance down to $Spin(7)$  \cite{bakasi}.}
supersymmetry in $8$ dimensions.

The eight-dimensional algebra has  been  built   \cite{BBT}  from the scalar and vector topological
symmetries  with   $9=1+8$ generators that can be algebraically constructed. These 9 generators 
build  a  maximally  closed and
consistent sector of the  twisted   $\mathcal{N}=2$, $d=8$
Yang--Mills supersymmetry. Invariance under this subalgebra completely determines the
Yang--Mills supersymmetric action. The full on-shell supersymmetry is recovered in this way and one can then interpret the invariance of the action under the $7$ additional supercharges as accidental.
\subsection{The $8$-dimensional BRSTQFT formula}
The nine $8$-dimensional supersymmetry generators can be encoded in a
graded differential operator $Q$ which 
depends on     nine twisted supersymmetry parameters consisting of one scalar
$\bar{\omega}$ and one eight-dimensional vector $\ve$. 

Using the notation of \cite{BBT},
$Q$  satisfies the horizontality condition in
eight dimensions
\bea\label{hor}
\hat F  _8\equiv (d+ Q-\bar{\omega}i_\varepsilon)(A+c)
+(A+c)^2\qquad\qquad\CR =
F+\bar{\omega}\psi+\d(\e)\eta+i_\varepsilon\chi +
\bar{\omega}^2\Phi + \vert\ve\vert^2\bar{\Phi} .
\eea
In Eq.~(\ref{hor}), all
fields are forms taking values in the Lie algebra  of the gauge group. 
For example, $A = A_i dx^i$ is the  Yang--Mills connection where $i=1$ to 8, 
$\Phi$ and $\bar\Phi$ are
scalars, and $F=dA + A A$ is a two-form. Furthermore, $\Psi
 = \Psi_i dx^i$ is a $1$-form,
$\chi  = \frac{1}{2} \chi_{ij} dx^i dx^j$ is an antiselfdual $2$-form   with seven
independent components, and $\eta$ is a scalar field where
$(\Psi_i , \chi_{ij}, \eta)$ are twisted Fermi spinors.  Moreover, 
$d$  is the usual  exterior differential $d=\partial_i  dx^i$, $i_v$
is the contraction operator  along the vector $v$, $i_\varepsilon
dx^i=    \varepsilon^i$, and  $\d(\ve)\equiv \ve^i\d_{ij}dx^j$.

Finally, the anticommuting scalar field $c$ is a shadow field. It   plays
an important role by  closing the supersymmetry  without  field-dependent 
gauge transformations in the right-hand-side of commutators,  and, eventually, for quantizing the theory
\cite{BBS} . There is no need at this stage to introduce a
Faddeev--Popov ghost. Note that all fields and operators have a grading that
is the sum of shadow  number and ordinary form degree.

The closure of $Q$ is ensured by the   Bianchi identity, which
also determines the action of the symmetry on the fields on the
right-hand-side of Eq.~(\ref{hor}). By expanding
the equation
\bea\label{bianchi}
(d+Q-\bar{\omega}i_\ve)(F+\bar{\omega}\psi+\d(\e)\eta+i_\varepsilon\chi +
\bar{\omega}^2\Phi + \vert\ve\vert^2 \bar{\Phi}) \nn\\ +
[A+c,\,F+\bar{\omega}\psi+\d(\e)\eta+i_\varepsilon\chi
+\bar{\omega}^2 \Phi + \vert\ve\vert^2 \bar{\Phi}]= 0 ,
\eea
one finds that the  action of the
operator $Q$
on the fields can be decomposed into
a gauge transformation with parameter $c$
and a supersymmetry transformation $\susy$ with 9 twisted
parameters $\bar{\omega}$ and  $\ve$ as
\be
Q=\susy-\gauge(c)=\bar{\omega}\delta_0+\ve^i \delta_i-\gauge(c) .
\ee
The  off-shell closure of $\susy$  follows from the identity $Q^2=
\bar{\omega}   \L_\varepsilon$.  Notice that no gauge
transformation is involved in this equation.

\subsection{Light-cone $10$-dimensional equation}

We may  understand Eq.~(\ref{hor}) as a light-cone projection of an
analogous equation in $10$ dimensions. In order to determine the
subalgebra of the 10-dimensional theory, we   ``oxidize" this
equation by introducing light-cone modes ($\del_+$,$\del_-$) and
redefining $(\Phi,\bar{\Phi})\to (A_+,A_-)$ in such a way that
\be\label{lightcone}
Spin(7)\times\mathds{R}_+^* \cong Spin(7)\times SO(1,1)
\subset SO(8)\times SO(1,1) \subset SO(1,9) .
\ee
In this way,  we can    interpret the scalar fields in  the
right-hand-side  of Eq.~(\ref{hor}) as elements of a connexion in
10 dimensions. They can be   carried to the
left-hand-side of   the    horizontality condition (\ref{hor}),
which thus  appears  as the dimensional reduction of the 
10-dimensional condition
\bea\label{hor10}
\hat  F_{10}\equiv (d+
Q-\bar{\omega}i_\varepsilon-\bar{\omega}^2i_+
-\vert\varepsilon\vert^2i_-)(A+c) +(A+c)^2\qquad\qquad\CR =
F+\bar{\omega}(\psi+\eta dx^-)+(\d(\e)\eta+i_\varepsilon\chi+i_\varepsilon\psi\,
dx^+).
\eea
Eq.~(\ref{hor10})  has the Bianchi identity
\bea
(d+ Q-\bar{\omega}i_\varepsilon-\bar{\omega}^2i_+
-\vert\varepsilon\vert^2i_-)(F+\bar{\omega}(\psi+\eta
dx^-)+(\d(\e)\eta+i_\varepsilon\chi+i_\varepsilon\psi\,
dx^+))\nn\\
+[A+c,F+\bar{\omega}(\psi+\eta
dx^-)+(\d(\e)\eta+i_\varepsilon\chi+i_\varepsilon\psi\,
dx^+)]=0,
\eea
which insures that $(d+ Q-\bar{\omega}i_\varepsilon-\bar{\omega}^2i_+
-\vert\varepsilon\vert^2i_-)^2=0$.


By expansion according to the  various gradings, we obtain
$10$-dimensional transformation laws for all fields, which exactly reproduce 
those described earlier and determined by a mere twist of supersymmetry transformations.
The  self-dual 2 form
auxiliary  fields  $G_{ij}$  with seven degrees of freedom is  now 
   introduced here in the standard TQFT way, by  solving   the
degenerate equations $\delta_0 \chi_{ij}    +\delta_i\Psi_j+\cdots=0$.  As a
consequence of the Bianchi identity,  the algebra of generators
$\delta_0$ and $\delta_i$  closes independently of any equations of
motion, as expressed in the preceding section.   Let us stress again the relevance of the shadow  $c$ for supressing field dependent gauge transformations in the commutators of supersymmetries.

The   $10$-dimensional  action (assuming no higher-derivative terms)
is completely determined from the $Q$-invariance with the nine parameters $\bar{\omega}$
and $\ve_i$. As in \cite{BBT}, one can show
that the most general $Q$-invariant
expression, which is independent of $\ve_i$ and contains no higher order derivative terms,  can be written either as a $\delta_0$-exact or as a
 $\delta_i$-exact functional up to a topological term 
\bea
 S&=&\delta_0{\bf Z}^{(-1)} - \frac{1}{8}\int_M d^{10} x\,
\trace \Scal{\Omega^{ijkl}F_{ij}F_{kl}} \CR &=& \ve^i\delta_i{\bf Z}^{(+1)} + \frac{1}{8}\int_M d^{10} x\,
\trace \Scal{\Omega^{ijkl}F_{ij}F_{kl}} \eea
where ${\bf Z}^{(-1)}$ and ${\bf Z}^{(+1)}$ are  completely fixed respectively by the
$\delta_i$ and $\delta_0$ symmetries, i.e. $\delta_i{\bf
Z}^{(-1)}=\delta_0{\bf Z}^{(+1)}=0$.

As will be shown in the following section, this matches the
Lagrangian obtained by twist in (\ref{10D action}). We have thus
obtained an off-shell formulation of ten-dimensional
super-Yang--Mills from the eight-dimensional Yang--Mills BRSTQFT.

\section{Toward a  superspace formulation in twisted fermionic variables}

The fact that we are able to obtain a subalgebra that closes without 
the use of the equations of motion suggests that there should
exist an off-shell superspace formulation of ten dimensional 
super-Yang--Mills. Since there are nine off-shell supersymmetry
generators, it is natural to define a superspace with nine
anticommuting variables. Let us define the reduced superspace with vector 
coordinates $\theta^i$ (spinor representations of $Spin(7)$) and scalar 
coordinate $\theta$. We define superspace derivatives
\begin{gather}    \hat{\nabla}=\frac{\partial \, }{\partial \theta} 
- \theta \partial_+ ,  
\hspace{10mm} \hat{\nabla}_i=\frac{\partial \, }{\partial \theta^i} - 
\theta \partial_i - \theta_i \partial_- , \CR
 \partial_+ ,\hspace{10mm} \partial_- ,\hspace{10mm} \partial_i, 
\end{gather}
which obey
\begin{gather}
\hat{\nabla}^2= -\partial_+ ,
 \hspace{10mm} \hat{\nabla}_{\{i}\hat{\nabla}_{j\}}= -\delta_{ij} \partial_-,
\hspace{10mm} \{\hat{\nabla},\hat{\nabla}_i\}= -\partial_i,
\end{gather}
with all other commutators equal to zero. For each of the 
superspace derivatives,
we introduce a corresponding gauge connection 
superfield and define the covariant superderivatives
\begin{gather}
\nabla=\hat{\nabla}+ \mathds{C},
 \hspace{10mm} \nabla_i=\hat{\nabla}_i+ {\Upgamma}_i, \CR
 \mathcal{D}_+=\partial_+ +\mathds{A}_+ , 
\hspace{7mm} \mathcal{D}_-=\partial_- +\mathds{A}_- ,
\hspace{7mm} \mathcal{D}_i =\partial_i+\mathds{A}_i.
\end{gather}

To reduce
the number of degrees of freedom, one needs to constrain the 
supercurvature associated to the connection superfields. The usual 
$SO(9,1)$-covariant constraint for ${\N = 1}$ $D=10$ super-Yang--Mills
superfields is 
$\{\nabla_{\alpha}, \nabla_{\beta}\} + 
2 \gamma_{\alpha\beta}^m \mathcal{D}_m =0,$ 
which puts the superfields on-shell. However, in the reduced superspace,
the analogous constraints are 
\begin{gather} \nabla^2 + \mathcal{D}_+ = 0, 
\hspace{10mm} \{ \nabla_i , \nabla_j \} + 2 \delta_{ij} {\cal D}_- = 0,\CR
 \{ \nabla , \nabla_i \} + {\cal D}_i = 0, \end{gather}

It is remarkable that the resolution of these constaints no longer imply the equations of motion, in contrast   with the case of the full superspace constraints, that can only be written on-shell.

As usual, the commutator of a fermionic  covariant derivative and a 
bosonic covariant derivative gives a fermionic gauge-covariant
superfield. The Bianchi identities imply that the symmetric part of $[\nabla_i, {\cal D}_j]$ is proportional to $\delta_{ij}$. In order to obtain the right supermultiplet we furthermore impose the constraint that the antisymmetric part of $[\nabla_i, {\cal D}_j]$ is antiselfdual
\be P_{ij}^{+kl} [\nabla_k, {\cal D}_l] = 0\ee
We therefore define the gauge-covariant superfields 
$\Uppsi_i,\, \upeta$ and $\upchi_{ij} = P_{ij}^{-kl} \upchi_{kl}$ 
that correspond to the 
commutators
 \begin{gather} [ \nabla , {\cal D}_i] \equiv \Uppsi_i
=-[ \nabla_i, {\cal D}_+],  \hspace{5mm}  
[ \nabla , {\cal D_-}] \equiv \upeta, \hspace{5mm} 
[\nabla , {\cal D}_+] =0, \nn \\ 
[\nabla_i, {\cal D}_-] =0,\hspace{5mm} [\nabla_i, {\cal D}_j] = -\d_{ij} \upeta - \upchi_{ij} .
\end{gather}

The constraints and their Bianchi identities imply that 
$\Uppsi$, $\upeta$ and $\upchi$ satisfy
\be \nabla_{\{i} \Uppsi_{j\}} + \delta_{ij} \nabla \upeta = 0,
 \hspace{10mm} \nabla_k \upchi_{ij} + 8\, {P^-_{ijk}}^l \nabla_l \upeta = 0. \ee
Furthermore, the
commutators of bosonic covariant derivative give the
superderivative of these fermionic superfields as
 \be\begin{split} [ {\cal D}_i , {\cal D}_-] &\equiv\mathds{F}_{i-}=
  \nabla_i \upeta ,\\  
[ {\cal D}_i , {\cal D}_+] &\equiv \mathds{F}_{i+}=
 \nabla \Uppsi_i ,\end{split} 
\hspace{10mm}\begin{split}
  [ {\cal D}_- , {\cal D}_+] &\equiv\mathds{F}_{-+}= \nabla \upeta,  \\  
[ {\cal D}_i , {\cal D}_j] &\equiv \mathds{F}_{ij}=
 \nabla \upchi_{ij} - \nabla_{[i} \Uppsi_{j]} .
\end{split}\ee
The superfields  $\mathds{C}$ and $ {\Upgamma}_i$ have expansion of the form
\be
\mathds{C}=c+\theta^i  c_i+\cdots \hspace{10mm} {\Upgamma}_i=\gamma_i  +\theta^j \gamma_{ij}+\cdots
\ee
The transformation laws are such that, $c$ can be identified as the shadow of   \cite{BBS} and 
$\gamma_i  \kappa^i$ as an analogous field introduced in \cite{BBT}, in the context of the topological vector symmetry, for $  \kappa$ a constant vector field.

In order to concretely realize the abstract algebra defined by the above equations,  one must   determine  the action of $\delta_0$ and $\delta_i$ on all components of $\mathds{C}$ and $ {\Upgamma}_i$,  which satisfy  the relevant commutation relations.  We have checked this non trivial property, both in component formalism  and directly in superfield formalism~\cite{TBP}, and prove thereby  that the above constraints  that hold off-shell can be solved and that the solution corresponds to the supermultiplet of ten dimensional super-Yang--Mills in its twisted formulation.

%
%
%

\subsection{Super-Yang--Mills action in superspace}

To write a superspace action in terms of these constrained
superfields, first note that the component action of (\ref{10D action})
can be written as a $\delta_0$-exact functional as long as we neglect instantons \bea
S=\delta_0{\bf Z}^{(-1)} \eea with ${\bf Z}^{(-1)}$ completely fixed
by the $\delta_i$ symmetry, i.e. $\delta_i{\bf Z}^{(-1)}=0$ where \bea
{\bf Z}^{(-1)}=\int_M d^{10} x\, \trace
\Scal{\frac{1}{2}\,\chi^{ij}(F_{ij}+\frac{1}{4}G_{ij})+F_{-i}\psi^i+\eta
F_{+-}} \eea Moreover, defining $\delta(\ve)=\ve^i\delta_i$, the
action can be expressed as a $\delta_0\delta(\ve)$-exact term as
\bea 
\label{exact}
S = \delta_0\delta(\ve)\int_M d^{10} x\,
\frac{1}{\vert\ve\vert^2}\mathscr{F} \eea with \bea \mathscr{F}=
\trace
\biggl(\frac{1}{4}\,\ve_i\Omega^{ijkl}(A_jF_{kl}-\frac{2}{3}A_jA_kA_l)
+\ve_i(-\delta^{ij}\eta-\chi^{ij})\psi_j
\biggr). \eea Note that $\mathscr{F}$ is completely constrained by
the condition that its $\delta(\ve)$ variation is independent of
$\ve$. 

This situation is reminiscent of the case of harmonic
superspace \cite{harmo,harmo10} where harmonic coordinates allow the construction
of manifestly supersymmetric actions using a reduced superspace.
In this case, one does not have harmonic variables but
one can nevertheless write the above action
in reduced superspace as 
\bea 
\label{supers}
S = \int_M d^{10} x\, \nabla \nabla_i \, {\bf K}^i \equiv
 \int_M d^{10} x\, \int d\theta d\theta_i \, {\bf K}^i 
\eea 
where
\bea {\bf K}^i =
\trace
\biggl(\frac{1}{4}\,\Omega^{ijkl}(\mathds{A}_j
\mathds{F}_{kl}-\frac{2}{3}\mathds{A}_j\mathds{A}_k\mathds{A}_l)
-(\delta^{ij}\upeta+\upchi^{ij})\Uppsi_j \biggr). 
\eea 
Since the 
$\delta(\ve)$ variation of ${1\over{|\ve|^2}}\mathscr{F}$ 
is independent of $\ve$, one learns that 
\bea
\label{Kident}
\nabla_i {\bf K}_j + \nabla_j {\bf K}_i = \d_{ij} f + 
{\partial\over{\partial x^\mu}}
h^\mu_{ij}
\eea
for some $f$ and $h^\mu_{ij}$. Using (\ref{Kident}), it is straightforward
to show that (\ref{supers}) is independent of $\t$ and $\t^i$
and is therefore invariant under all nine supersymmetries.

\subsection*{Acknowledgments}

This work was partially supported under the contract ANR(CNRS-USAR) \\ \texttt{no.05-BLAN-0079-01}.


\begin{thebibliography}{50}



\bibitem{berko}
 N.~Berkovits, ``A ten-dimensional super-Yang--Mills action with
off-shell supersymmetry'', Phys.\ Lett.\     B   {\bf 318} (1993) 104,
\texttt{[arXiv:hep-th/9308128]}.


 \bibitem{BBT}
 L.~Baulieu, G.~Bossard and A.~Tanzini,
 ``Topological vector symmetry of BRSTQFT and construction of maximal
 supersymmetry,''
 JHEP\ {\bf 0508} (2005) 037
 \texttt{[arXiv:hep-th/0504224]}.

  \bibitem{BBS}
     L.~Baulieu, G.~Bossard and S.~Sorella,
  ``Shadow fields and local supersymmetric gauges,''
Nucl.\ Phys.\ B  {\bf 753}   (2006) 252,
\texttt{[arXiv:hep-th/0603248]}.

\bibitem{Evans}
  J.~M.~Evans,
``Supersymmetry algebras and Lorentz invariance for d = 10 superYang--Mills,''
  Phys.\ Lett.\ B {\bf 334} (1994) 105.

\bibitem{harmo}
  A.~S.~Galperin, E.~A.~Ivanov, V.~I.~Ogievetsky and E.~S.~Sokatchev,
  ``Harmonic Superspace,'' Cambridge, UK: Univ.\ Pr.\ (2001) 306 p.

\bibitem{harmo10}
 E.~Nissimov, S.~Pacheva and S.~Solomon, 
``Off-shell superspace d=10 super-Yang--Mills from covariantly quantized Green--Schwarz superstring,''
Nucl.\ Phys.\ B {\bf 317} (1989) 344;\\
E.~Nissimov, S.~Pacheva and S.~Solomon,
``Action principle for overdetermined systems of nonlinear field equations,''
Int.\ J.\ Mod.\ Phys.\ A {\bf 4} (1989) 737.



\bibitem{bakasi} 
L.~Baulieu,  H.~Kanno, I.~M.~Singer,  
``Special quantum field theories in eight and other
dimensions,''
Commun.\  Math.\  Phys.\   {\bf  194} (1998) 149, 
\texttt{[arXiv:hep-th/9704167]};\\
L.~Baulieu and P.~West,
``Six-dimensional TQFT's and twisted supersymmetry,''
Phys.\ Lett.\      B  {\bf 436} (1998) 97,
\texttt{[arXiv:hep-th/9805200]}.  
  
  
 \bibitem{TBP}
  {  In preparation}.

 
\end{thebibliography}
\end{document}